\documentstyle[12pt,aaspp4]{article}
\begin{document}

\title{The Effect of Environment on the X-Ray Emission from Early-Type Galaxies } 

\author{Beth A. Brown \altaffilmark{1} and Joel N. Bregman} 
\affil{Department of Astronomy, University of Michigan, Ann Arbor, MI 48109-1090\\
Beth.Brown@gsfc.nasa.gov, jbregman@umich.edu}
\altaffiltext{1}{National Research Council Research Associate at 
NASA/Goddard Space Flight Center, Greenbelt, MD}

\begin{abstract}
In order to help understand the phenomena of X-ray emission from early-type 
galaxies, we obtained an optically flux-limited sample of 34 elliptical and S0 
galaxies, observed with high and low angular resolution detectors on the X-ray 
R\"{o}ntgen Satellite ({\em ROSAT}).  Our analysis of this sample, discussed 
previously, suggested that the most X-ray luminous galaxies were in rich
environments.  Here we investigate environmental influences quantitatively 
and find a positive correlation between $L_B/L_X$ and the local galaxy 
density.  Since the local galaxy density is usually related to the density of hot
intergalactic gas, we suggest that this correlation occurs because the
X-ray luminosity is enhanced either through accretion of the intergalactic 
gas or because the ambient medium stifles galactic winds.  When the ambient 
medium is 
unimportant, partial or global galactic winds can occur, reducing $L_B/L_X$.
These effects lead to the large observed dispersion in $L_X$ at fixed $L_B$.
The transition from global winds to partial winds occurs in the $L_B$ range
of our sample and we argue that this transition is one of the 
principle reasons for the steep relationship between $L_X$ and $L_B$.
This is a significant departure from the steady-state cooling flow
interpretation and we discuss predictions that can be tested with future 
observatories.

We discuss details of the data reduction not previously presented, and
examine the dependence of $L_X$ on the choice of outer source radius 
and background location.  Effects of Malmquist bias are shown not to be
important for the issues addressed.  Finally, we compare the temperature
deduced for these galaxies from different analyses of {\em ROSAT} and 
{\em ASCA} data.  The {\em ASCA} and {\em ROSAT} temperatures are similar, 
provided that the same metallicities and Galactic absorption columns are used.  
However, temperatures have a dependence on the metallicity and 
absorption column, especially for the lower temperature systems.
\end{abstract}


\section{Introduction}
\label{sec:intro} 

There are two primary sources for the X-ray luminosity of early-type galaxies,
emission from stellar sources and optically thin radiation from a hot dilute
interstellar medium (e.g., Fabbiano \markcite{fao89}1989). For the more luminous 
X-ray emitting ellipticals, there is little doubt that emission from hot interstellar gas 
is the dominant mechanism. To explain the behavior of the X-ray emission
from hot gas, cooling flow models were developed, and had a number of
successes. In the standard model, gas that is shed by stars is converted into hot gas 
with a temperature corresponding to the velocity dispersion of the stars. Radiative 
losses of the gas cause it to fall inward into the galaxy, releasing additional
gravitational energy. This model is able to account for the observed magnitude
of the X-ray luminosity, it can reproduce the X-ray surface brightness
distribution, and it predicts a correlation between $L_X$ and $L_B$. However, 
there are a few important discrepancies between theory and observation. The
predicted slope of $L_X$--$L_B$ relationship (approximately $L_X \propto L_B^m$,
where $m=1.6$--1.8) is not as steep as the observational relationship obtained
using {\em Einstein Observatory} data, where $m=1.7$--2.4. More recent work by 
Brown \& Bregman (\markcite{bro98}1998) and Irwin \& Sarazin
(\markcite{irw98}1998) find the slope to be about 2.7, from {\em ROSAT} data, 
in clear disagreement with the steady-state cooling flow model. Another 
important discrepancy is the extremely large dispersion about the relationship, 
about two orders of magnitude in $L_X$ at fixed $L_B$, first discovered in  
{\em Einstein Observatory} data (Canizares, Fabbiano, \& Trinchieri 
\markcite{can87}1987) and confirmed with the {\em ROSAT} 
sample (White \& Davis \markcite{whi97}1997; Brown \& Bregman \markcite{bro98}1998).

There have been a few explanations for the cause of the large dispersion in
$L_X$, which is either attributed to the internal properties of the galaxy or 
to the external influences of the environment. D'Ercole et al. 
(\markcite{der89}1989) suggested
that slight differences in the structure of a galaxy could lead to widely
variant X-ray luminosities at fixed optical luminosity. However, this result
occurs for a particular form of the supernova rate as a function of cosmic
time, which may not be likely (Loewenstein \& Mathews \markcite{loe91}1991). Other models that include
the effects of supernova heating and galactic winds do not predict a large
dispersion in $L_X$ at the current epoch 
(David, Forman, \& Jones \markcite{dav91}1991; 
Loewenstein \& Mathews \markcite{loe87}1987; Vedder, Trester, \& Canizares \markcite{ved88}1988).

Environmental effects are becoming more widely understood to have a
significant, if not central influence on the X-ray properties of a galaxy.
One effect is stripping of gas from a galaxy as it passes through the ambient
cluster medium (Takeda, Nulsen, \& Fabian \markcite{tak84}1984; Gaetz, Salpeter, \& Shaviv \markcite{gae87}1987). This is likely to be important in the richer 
clusters, but most current samples do not contain very rich clusters
(the Virgo Cluster being the richest). Although there is evidence for
stripping for NGC 4406 in Virgo, early-type galaxies in Virgo are not X-ray
underluminous in general, so stripping is probably the exception rather than
the rule. Another environmental effect is accretion of material onto galaxies,
which can raise the X-ray luminosity substantially (Brighenti \& Mathews \markcite{bri98}1998) and probably is most important in gas-rich group and 
cluster environments. A third environmental effect is the ability of an ambient 
medium to stifle a galactic wind, causing those galaxies to retain their gas 
and become more X-ray luminous than field galaxies (Brown \& Bregman \markcite{bro98}1998).

In Brown \& Bregman (\markcite{bro98}1998), we presented the X-ray 
luminosities of a complete optically-selected sample of 34 early-type galaxies
in the 0.5--2.0 keV {\em ROSAT} band. We noticed that the galaxies with the 
largest values of $L_X$ (for a given $L_B$) were in the more populated regions (the 
Virgo or Fornax clusters, or in the centers of groups) of the sample, suggesting 
that being in moderately rich environments enhances the X-ray emission of a 
system. This trend may differ from that reported by White \& Sarazin 
(\markcite{whi91}1991), who found that lower-X-ray luminous galaxies had
50\% more bright neighboring galaxies than higher-X-ray luminous galaxies. To 
further assess the importance of environment upon X-ray luminosities, we 
advance our previous work by analyzing the X-ray properties of our sample in 
the context of a quantitative measure of the galactic richness provided by 
Tully (\markcite{tul88}1988). In addition, we present information on data 
processing, sample bias, temperature distribution, and several other issues not 
discussed in our previous Letter.

\section{Galaxy Sample and Data Processing}
\label{sec:sample}

The criteria for including a galaxy in this work is established in 
Brown \& Bregman (\markcite{bro98}1998). The targets (see
Table \ref{tab:opt}) are detected at the $\geq$ 97\% confidence
level, and include twelve galaxies lying within the poor clusters of 
Fornax and Virgo, with the remainder either isolated or lying in loose groups.  
For the purpose of this survey, we used PSPC data in the processing rather than 
HRI data when available, since HRI data contains no spectral information
(Table \ref{tab:obs}).

{\small
\begin{table}[p]
\vspace{-0.3truein}
\caption{Optical Galaxy Properties \label{tab:opt}}
\begin{center}
\newcommand{\nb}{\phm{\tablenotemark{b}}}
\newcommand{\nc}{\phm{\tablenotemark{c}}}
\newcommand{\erra}[1]{$\scriptstyle \pm{#1}$}

%\begin{tabular}{l@{\hspace{-3mm}}crrrcrrr@{\hspace{-0.05mm}}r}
\begin{tabular}{lcrrrcrrrr}
\tableline \tableline

Name & Type & {\em l \hspace{4mm}} & {\em b \hspace{3mm}} & $B^0_T$ 
\hspace{2mm} &
log $\sigma$ & D \hspace{5mm} & $r_e$ \hspace{2mm} &
log$L_B$ \hspace{2mm} & $\rho$ \hspace{2mm} \\
{} &  & \multicolumn{2}{c}{(deg)} & (mag) &
(km/s) & (km/s) \hspace{1mm} & ($\arcsec$)\hspace{2mm} &
(ergs/s) \hspace{2mm} & Mpc$^{-3}$ \\ \tableline

N 0720 &E &  173.02 & -70.36 & 11.16 & 2.392 & 2050 \erra{435}\nb &  38.58\nc &  43.65 \erra{0.06} & 0.25 \\
N 1316\tablenotemark{a}& S0 &  240.16 & -56.69 & 9.40 & 2.401 & 1422 \erra{\phn 88}\nb &  80.14\nc &  44.04 \erra{0.06} & 1.15 \\
N 1344 &E &  229.07 & -55.68 & 11.11 & 2.204 & 1422 \erra{\phn 88}\nb  &  38.38\nc &  43.35 \erra{0.06} & 0.28 \\
N 1395 &E &  216.21 & -52.12 & 10.94 & 2.412 & 1990 \erra{187}\nb &  45.07\nc &  43.71 \erra{0.06} & 0.55 \\
N 1399 &E &  236.72 & -53.63 & 10.55 & 2.491 & 1422 \erra{\phn 88}\nb  &  42.37\nc &  43.58 \erra{0.06} & 1.59 \\
N 1404 &E &  236.95 & -53.56 & 10.89 & 2.353 & 1422 \erra{\phn 88}\nb  &  26.72\nc &  43.44 \erra{0.06} & 1.59 \\
N 1407 &E &  209.64 & -50.38 & 10.57 & 2.455 & 1990 \erra{187}\nb &  71.96\nc &  43.86 \erra{0.12} & 0.42 \\
N 1549 &E &  265.41 & -43.80 & 10.58 & 2.312 & 1213 \erra{256}\nb &  47.44\nc &  43.43 \erra{0.06} & 0.97 \\
N 2768 &E &  155.49 &  40.56 & 10.93 & 2.296 & 1532 \erra{325}\nb &  49.44\nc &  43.49 \erra{0.12} & 0.31 \\
N 3115 &S0 &  247.78 & 36.78 &  9.95 & 2.425 & 1021 \erra{215}\nb &  32.32\nc &  43.53 \erra{0.06} & 0.08 \\
N 3377 &E &  231.18 &  58.31 & 11.13 & 2.116 &  857 \erra{126}\nb &  32.75\nc &  42.91 \erra{0.12} & 0.49 \\
N 3379 &E &  233.49 &  57.63 & 10.43 & 2.303 &  857 \erra{126}\nb &  35.19\nc &  43.19 \erra{0.06} & 0.52 \\
N 3557 &E &  281.58 &  21.09 & 11.13 & 2.465 & 2399 \erra{509}\nb &  37.10\nc &  43.80 \erra{0.06} & 0.28 \\
N 3585 &E &  277.25 &  31.17 & 10.53 & 2.343 & 1177 \erra{249}\nb &  38.04\nc &  43.42 \erra{0.06} & 0.12 \\
N 3607 &S0 &  230.59 & 66.42 & 10.53 & 2.394 & 1991 \erra{242}\nb &  65.49\nc &  43.88 \erra{0.12} & 0.34 \\
N 3923 &E &  287.28 &  32.22 & 10.52 & 2.335 & 1583 \erra{236}\nb &  52.16\nc &  43.68 \erra{0.06} & 0.40 \\
N 4125 &E &  130.19 &  51.34 & 10.58 & 2.359 & 1986 \erra{295}\nb &  58.39\nc &  43.86 \erra{0.12} & 0.34 \\
N 4278 &E &  193.78 &  82.77 & 11.02 & 2.425 & 1470 \erra{218}\nb &  32.89\nc &  43.42 \erra{0.06} & 1.25 \\
N 4365 &E &  283.80 &  69.18 & 10.64 & 2.394 & 1333 \erra{\phn 71}\nb  &  56.57\nc &  43.48 \erra{0.06} & 2.93 \\
N 4374 &E &  278.20 &  74.48 & 10.13 & 2.458 & 1333 \erra{\phn 71}\nb  &  54.47\nc &  43.69 \erra{0.06} & 3.99 \\
N 4406 &E &  279.08 &  74.64 &  9.87 & 2.398 & 1333 \erra{\phn 71}\nb  &  89.65\nc &  43.79 \erra{0.06} & 1.41 \\
N 4472 &E &  286.92 &  70.19 &  9.32 & 2.458 & 1333 \erra{\phn 71}\nb  & 103.61\nc &  44.01 \erra{0.06} & 3.31 \\
N 4494 &E &  228.62 &  85.32 & 10.69 & 2.095 &  695 \erra{147}\nb &  45.13\nc &  42.90 \erra{0.06} & 1.04 \\
N 4552 &E &  287.93 &  74.97 & 10.84 & 2.417 & 1333 \erra{\phn 71}\nb  &  30.00\nc &  43.40 \erra{0.06} & 2.97 \\
N 4621 &E &  294.37 &  74.36 & 10.65 & 2.381 & 1333 \erra{\phn 71}\nb  &  45.54\nc &  43.48 \erra{0.12} & 2.60 \\
N 4636 &E &  297.75 &  65.47 & 10.20 & 2.281 & 1333 \erra{\phn 71}\nb  & 101.14\nc &  43.66 \erra{0.06} & 1.33 \\
N 4649 &E &  295.87 &  74.32 &  9.77 & 2.533 & 1333 \erra{\phn 71}\nb  &  73.42\nc &  43.83 \erra{0.06} & 3.49 \\
N 4697 &E &  301.63 &  57.06 & 10.03 & 2.218 &  794 \erra{168}\nb &  73.51\nc &  43.28 \erra{0.06} & 0.60 \\
N 5061 &E &  310.25 &  35.66 & 11.06 & 2.282 & 1196 \erra{253}\nb &  25.51\nc &  43.22 \erra{0.06} & 0.31 \\
N 5102 &S0 &  309.73 & 25.84 & 10.57 & 1.820 &  155 \erra{\phn 15}\tablenotemark{b} &  23.29\nc &  41.64 \erra{0.12} & 0.17 \\
N 5322 &E &  110.28 &  55.49 & 11.09 & 2.350 & 1661 \erra{352}\nb &  34.76\nc &  43.50 \erra{0.12} & 0.43 \\
N 5846 &E &    0.43 &  48.80 & 10.67 & 2.444 & 2336 \erra{284}\nb &  82.61\nc &  43.96 \erra{0.12} & 0.84 \\
I 1459 &E &    4.66 & -64.11 & 10.88 & 2.488 & 2225 \erra{472}\nb &  34.03\tablenotemark{c} &  43.83 \erra{0.06} & 0.28 \\
N 7507 &E &   23.44 & -68.04 & 11.15 & 2.377 & 1750 \erra{371}\nb &  31.41\nc & 43.52 \erra{0.06} & 0.09 \\ \tableline

\end{tabular}
\end{center}
%\tablenotetext{}{ 
{\footnotesize
Hubble galaxy type (col 2) and local galaxy densities
(column 10) from Tully (1988). 
Galactic coordinates (cols 3-4) from NED. Total $B^0_T$ magnitudes, velocity dispersions, and distances 
(cols 5-7) from Faber et al. (1989). Effective radii and blue optical luminosities (cols 8-9) derived
from Faber et al. (1989) values.\\
%\tablenotetext{a}{
$^a$Velocity dispersion from Roberts et al. (1991) adopted. Fornax
cluster distance adopted for galaxy distance. \\
%\tablenotetext{b}{
$^b$Distance for NGC 5102 from McMillan et al. (1994).\\
%\tablenotetext{c}{
$^c$The effective radius, $r_e$, for IC 1459 adopted from RC3 catalog.}
\end{table}
}

{\small
\begin{table}[p]
\begin{center}
\caption{Galaxy Observations \label{tab:obs}}
%\tablewidth{6in}

%\begin{tabular}{lllrrr}
\begin{tabular}{l@{\hspace{0.35in}}l@{\hspace{0.35in}}l@{\hspace{0.35in}}r@{\hspace{0.35in}}r@{\hspace{0.35in}}r}
\tableline \tableline
{} &  &  &  & Photon & \\
Name & Detector & Sequence no. & Livetime & Counts & $\Delta$Counts \\
{} &  &  & (sec) & \multicolumn{2}{c}{total no. in 4$r_e$} \\ \tableline

N 0720  & PSPC & rp600005n00 & 22169 &  1698.5 &  55.2 \\
N 1316  & PSPC & rp700437n00 & 23067 &  3277.2 &  92.1 \\
N 1344\tablenotemark{a} & PSPC & rp600529 & 4947 & 25.4 &  9.2 \\  
N 1395  & PSPC & rp600133n00 & 19754 &  1373.4 &  53.2 \\
N 1399  & PSPC & rp600043n00 & 52009 & 18009.2 & 186.4 \\
N 1404  & PSPC & rp600043n00 & 52009 & 13408.9 & 131.6 \\
N 1407  & PSPC & rp600163n00 & 20971 &  2388.8 &  71.0 \\
N 1549  & PSPC & wp600623n00 & 15320 &   477.1 &  34.8 \\
N 2768  & PSPC & rp60052700  &  4771 &   136.5 &  18.6 \\
N 3115  & PSPC & rp600120n00 &  7452 &   121.9 &  18.0 \\
N 3377\tablenotemark{a} & HRI & rh600830n00 & 33876 & 87.4 & 25.5 \\
N 3379  & HRI  & rh600829n00 & 24240 &   169.7 &  67.6 \\
N 3557  & PSPC & wp600464a01 & 19628 &   261.4 &  29.7 \\
N 3585  & PSPC & rp60052400  &  5199 &    49.5 &  11.8 \\
N 3607  & PSPC & rp600263n00 & 23865 &  1267.1 &  63.4 \\
N 3923  & PSPC & rp600533n00 & 22377 &  1456.1 &  49.7 \\
N 4125  & PSPC & rp600253n00 &  5709 &   434.3 &  30.3 \\ 
N 4278  & PSPC & rp701413n00 &  3413 &   170.8 &  17.3 \\
N 4365  & PSPC & rp600009n00 &  4213 &   632.4 &  47.6 \\
N 4374  & HRI  & rh600493n00 & 26237 &  1609.9 &  89.1 \\
N 4406  & PSPC & wp600105    & 22141 & 19590.6 & 203.5 \\
N 4472  & PSPC & rp600248n00 & 25971 & 21015.7 & 191.7 \\
N 4494  & PSPC & rp600162n00 & 11638 &    98.6 &  39.0 \\
N 4552  & PSPC & wp600586n00 & 16671 &  1811.2 &  51.0 \\
N 4621  & PSPC & rp600017n00 & 13096 &   164.7 &  34.1 \\
N 4636  & PSPC & rp600016n00 & 12100 & 10943.4 & 139.2 \\
N 4649  & PSPC & rp600017n00 & 13096 &  5423.8 &  93.7 \\
N 4697  & PSPC & rp600262a02 & 45259 &  2361.7 &  98.8 \\
N 5061\tablenotemark{a} & PSPC & rp600528 & 6516 & 28.6 & 8.9 \\
N 5102\tablenotemark{a} & PSPC & rp60052600 & 8937 & 36.7 & 11.7 \\
N 5322  & PSPC & rp600270n00 & 31556 &   502.6 &  43.4 \\
N 5846  & PSPC & rp600257    &  8808 &  3653.3 &  77.7 \\
I 1459  & PSPC & rp600266n00 & 32707 &  2605.4 &  59.6 \\
N 7507\tablenotemark{a} & HRI & rh600684n00 &  5469 & 22.4 & 12.0 \\ \tableline
\end{tabular}
\end{center}
%\tablenotetext{a}{
$^a$Total photon count tabulated within 1$r_e$ and extrapolated to 4$r_e$.
\end{table}
}

Data analysis was performed using the PROS system under NOAO's IRAF 
(software written specifically for X-ray data) with an objective of obtaining a 
well-defined $L_X$ for each galaxy in the sample (\S\ref{sec:source}), and 
$T_X$ (X-ray gas temperature) when possible.  Spatial analyses 
(\S\ref{sec:spatial}) were performed on images using a blocking factor of eight 
for the normalized PSPC data (neglecting the softest energy channels,
PI $<$ 20), creating 4 arcsec pixels.  Blocking factors of two or four were
used for HRI data resulting in 1-2 arcsec pixels. Spectral analyses
(\S\ref{sec:spectral}) were performed on normalized QPOE files using the same 
regions chosen for the spatial analyses, and binned according to the quality of 
the data. 

\subsection{Source and Background}
\label{sec:source}

The extent of the region within which the X-ray flux is determined (the
source), and the image regions chosen for background subtraction, can
have a significant effect on the determination of $L_X$ because the galaxies 
in our sample may sit in clusters or loose groups. Brown \& Bregman 
(\markcite{bro98}1998) briefly state the source and background radial
limits, and justifies the use of de Vaucouleur's half-light radius, $r_e$, as
the basis for those limits.  We chose to use a 4$r_e$ radius for the advantages 
it offers. A circle of 4$r_e$ is large enough to be resolved by the {\em ROSAT} 
PSPC for each galaxy, yet small enough to avoid possible cluster emission problems. 
For our five weakest targets (NGC 1344, NGC 3377, NGC 5061, NGC 5102, 
and NGC 7507) background emission dominates the signal when going out to 
4$r_e$. We therefore obtained the flux within 1$r_e$, to get its best measure while maximizing the S/N. That value is then multiplied by a factor of 1.58 to
extrapolate the flux out to 4$r_e$. The correction factor is obtained by taking the
ratio of a beta model ($\beta$=0.5, $r_{core} = r_e / 11$)  integrated from  
0 to 4$r_e$ to one integrated from 0 to 1$r_e$.

The background is chosen based on our desire to examine $L_X$ only within
4$r_e$.  If the background is defined to be a region far from the source
center, as is usually done, there may still exist cluster emission past
4$r_e$ surrounding the source.  To establish a flux within 4$r_e$, the excess
emission along our line of sight must be removed.  The best way to
accomplish this is to subtract an annulus within which the mean surface 
brightness equals the mean surface brightness of unwanted emission in
front of and behind the source.  We mathematically determined the radial 
boundaries of this annulus, using a beta model for extended emission in 
elliptical galaxies of the form
\begin{equation}
I_X = I_o[ 1 + (r / r_{core,X})^2]^{-3\beta + 0.5}
\end{equation}
where $I_o$ is the central surface brightness, and $r_{core,X}$ is the core radius
of the X-ray emission.
Although an infinite range of inner and outer radii is possible, we find that, 
to maximize the signal-to-noise, a unique solution of $r_1 = 4r_e$ and $r_2 = 
6.3r_e$ exist for $\beta = 0.5$.

\subsubsection{NGC 4494}

One galaxy in the sample, NGC 4494, lies far off-axis ($\sim$45$^{\arcmin}$
off). At this radius, the PSF becomes large and distorted, so we needed to 
correct for the difference between the on-center position and its off-axis 
position.  We could not simply scale the amount within 1$r_e$ to 4$r_e$ because 
1$r_e$ is significantly smaller than the PSF of the instrument, so the 
instrument scatters most of the photons out of 1$r_e$.  This is the only object 
that we have where that is true.  Instead, we chose a circle of radius 
240$^{\arcsec}$ as the source, and similar sized circles on either side as the 
background.  As a comparison, we calculated the flux using an annular background 4--6.3 $r_e$ from the center of the source which resulted in only a 10\% difference, 
smaller than the uncertainty due to photon statistics.

\subsection{Spatial Analysis}
\label{sec:spatial}

\begin{figure}
\begin{center}
%\figurenum{1}
\plotone{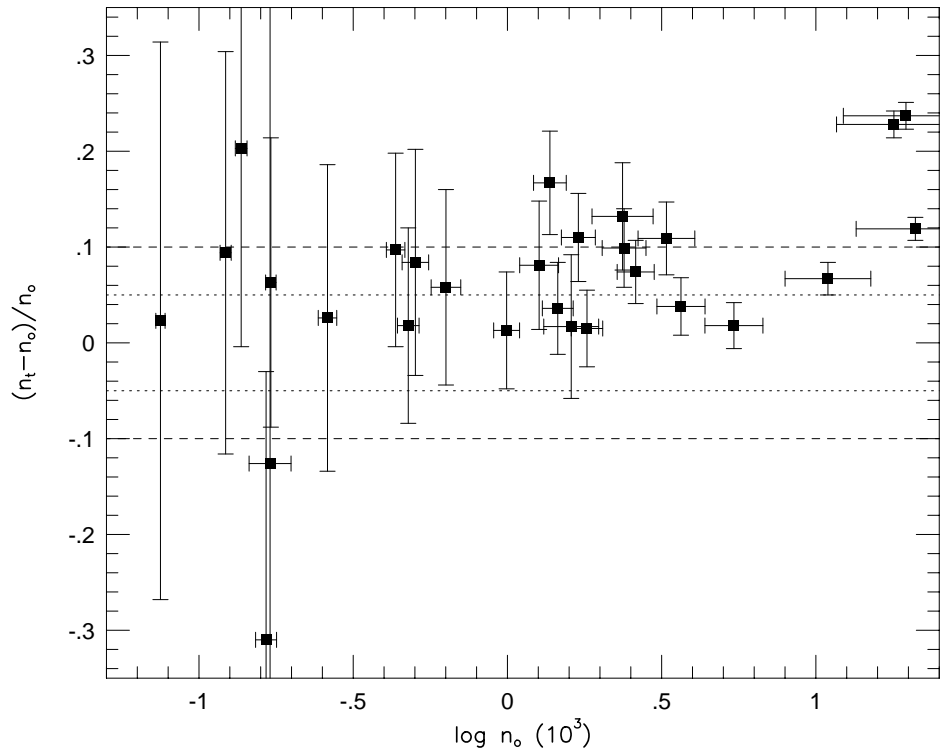}
\caption{The percent 
difference in the number of background-subtracted photon counts within a 
radius of 4$r_e$ using a background at 4-6.3$r_e$ (n$_o$) and a background 
taken at large radii (n$_t$) against log n$_o$. The dotted and dashed lines are 
at 5\% and 10\% respectively.
\label{fig:backg}}
\end{center}
\end{figure}

Spatial analysis was performed on the data in order to compare our well defined 
choice of background with backgrounds taken at large radii.  We first determined 
the total photon count (proportional to the flux) in a $4r_e$ circle 
using the 4--6.3 $r_e$ background (see Table \ref{tab:obs}). This flux was then 
compared to one obtained by subtracting a background region beyond the extent of 
the X-ray emission. This background region was also taken to be an annulus, and was typically located at $> 7 r_e$ from the galaxy center. The comparisons show that only for 36\% of the galaxies is 
there a difference greater than 10\% between taking a background as detailed 
above, and a more traditional background (neglecting the 
five weak detections and NGC 4494). We find that 
differences greater than 10\% only exist for four of the galaxies (NGC 1395, 
NGC 1399, NGC 4406, and NGC 4472) when photon count errors are taken into 
account, and 68\% of the galaxies show differences of less than 5\% (Figure 
\ref{fig:backg}).

\subsection{Spectral Analysis}
\label{sec:spectral}

We used a Raymond-Smith plasma model for spectral fits, succinctly
discussed in Brown \& Bregman (\markcite{bro98}1998).  Abundances were 
held at 0.5 solar since the reliability with which they can be determined from the 
{\em ROSAT} PSPC has been questioned (Bauer \& Bregman 
\markcite{bau96}1996).  A recent paper by Loewenstein \& Mushotzky 
(\markcite{loe97}1997) indicate that the metallicity for the galaxies that they 
observed with {\em {\em ASCA}} is about 0.5 solar, with scatter, making our 
choice a reasonable one. 
Single-temperature models were preferred over two-temperature models
when acceptable $\chi_{\nu} ^2$ values could be obtained ($\chi _{\nu}^2 > 1.46$ 
for 30 degrees of freedom).  For NGC 4472, an acceptable fit could 
not be obtained with either a single-temperature or two-temperature model with 50\% 
abundances, so a single-temperature model with 0.8 solar abundance was used.  
For NGC 1399 and NGC 1404, we also allowed the Galactic $N_H$ column density to 
be a free parameter within a limited range (20.1 $\leq$ log $N_H \leq$ 20.5 and
20.1 $\leq$ log $N_H \leq$ 20.3, respectively in $cm^{-2}$) to obtain an acceptable fit.  

Temperatures cannot be accurately fitted for low-count (i.e., $< 300$ counts)
PSPC objects, or HRI galaxies (no spectral information).  However, a fit
must be performed to the data to obtain a flux using PROS software.
Therefore, the data for these few galaxies were rebinned into single
bins, and fitted for the normalization only.  We assumed a fixed 
$T_X$ = 3/2$T_{\sigma}$ (typical of the findings of Davis \& White 
\markcite{dav96}1996) except for NGC 5102 and NGC 4494 where we used 
temperatures of 0.5 keV and 0.3 keV respectively, since the lower calculated 
temperatures led to clearly unacceptable fits. The stellar velocity dispersion 
temperature, $T_{\sigma}$, is calculated according to
\begin{equation}
kT = \mu m_p \sigma ^2,
\end{equation}
where $\mu$ is the mean molecular weight, and $\sigma$ is the one-dimensional 
stellar velocity dispersion.

After fitting the data, fluxes were obtained for two energy ranges $-$ 
0.5--2.0 keV and 0.1--2.0 keV $-$ using the 
intrinsic spectrum found for each source, and fitted temperatures when 
available.  Luminosities were then calculated using distances derived 
from Faber et al. (\markcite{fab89}1989) and an $H_0$ = 50 km/s/Mpc, 
with the exception of NGC 5102 for which no distance was given in the
Faber et al. (\markcite{fab89}1989) catalog (the McMillan, 
Ciardullo, \& Jacoby \markcite{mcm94}1994 distance is used for NGC 5102).  In the past, 
using these distances has significantly reduced the dispersion about $L_X$--$L_B$ 
line by increasing the internal consistency of distances (Donnelly, Faber, \& 
O'Connell \markcite{don90}1990).  Errors in the luminosity, due to uncertainties 
in the Galactic $N_H$ column density (typically 5--10\%) and uncertainties in photon counts, 
were examined and calculated. In all cases, errors in the photon statistics were 
found to be significantly greater than errors introduced from the uncertainty in 
the Galactic $N_H$ column.

\section{Analysis of the Observational Results}

We have reported briefly our findings of $L_X$ and $T_X$ for this study
in Brown \& Bregman (\markcite{bro98}1998).  We found a slope for the 
$L_X$--$L_B$ relationship steeper than previous investigations with 
{\em Einstein Observatory} data, with broad dispersion about the fit line.  
The observed $L_X$ of the brightest galaxies were found to be comparable to 
either the energy released through supernovae or through gravitational infall. 
A correlation was confirmed between $T_X$ and $T_{\sigma}$, using the 19 
high-photon-count PSPC galaxies, with a slope steeper than that reported by Davis 
\& White (\markcite{dav96}1996). We discussed a possible connection between the 
X-ray luminous galaxies and their gas temperatures with respect to 
$T_{\sigma}$, and suggested a possible connection between the observed X-ray 
luminosity and the environment in which each galaxy lies. Here, we present a 
more detailed analysis of the $L_X$--$L_B$ correlation (\S\ref{sec:lxlb}), the 
relationship between $T_X$ and $T_{\sigma}$ (\S\ref{sec:txtsig}), and the role 
environment may play in the observed luminosities (\S\ref{sec:environment}).

\subsection{The $L_X$--$L_B$ Plane}
\label{sec:lxlb}

\begin{figure}
%\figurenum{2}
\plotone{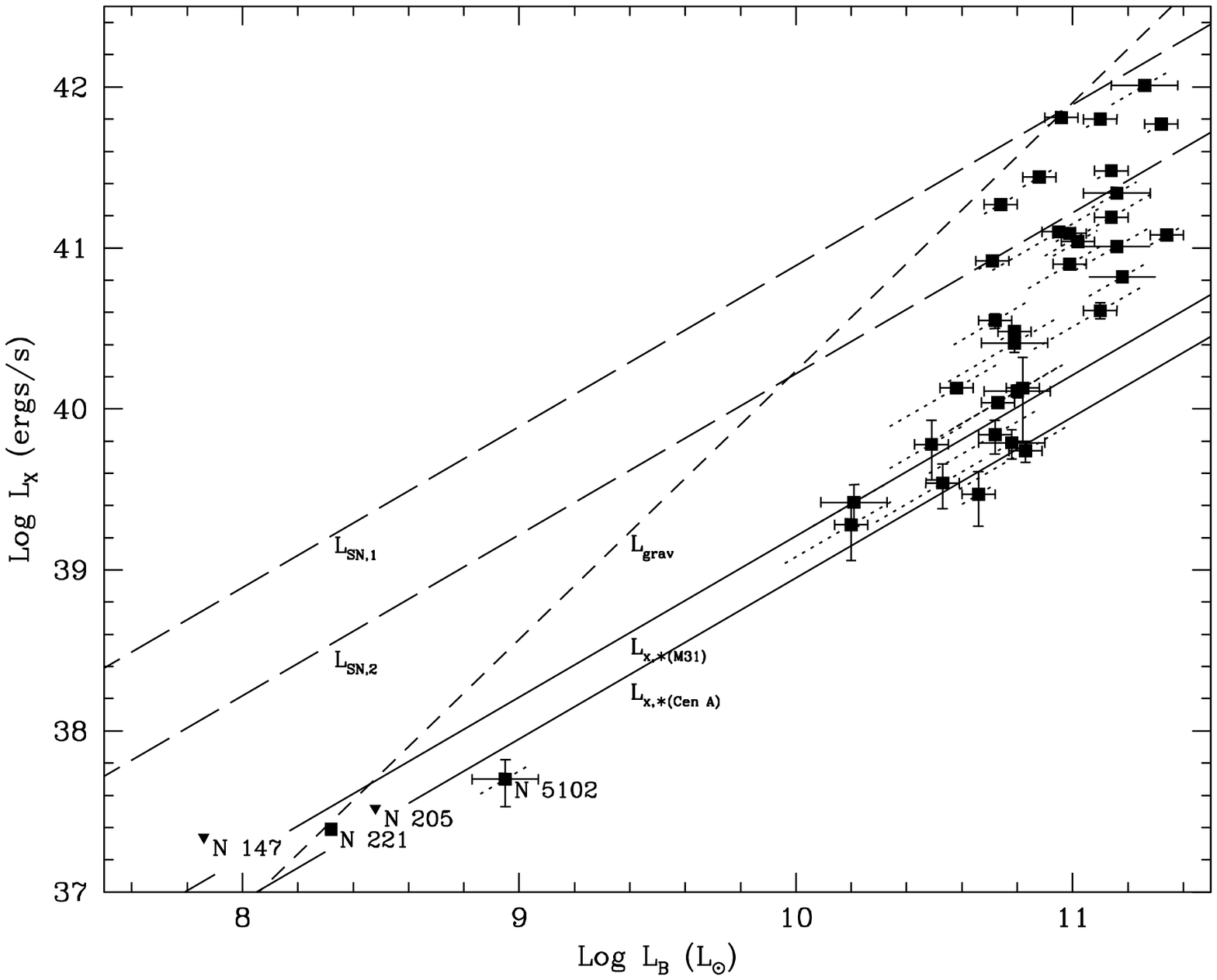}
\caption{ The optical blue luminosities ($L_B$) plotted
against the $0.5-2.0$ keV X-ray luminosities ($L_X$) from the {\em ROSAT}
PSPC and HRI instruments.  In addition to our sample, the dwarf galaxies NGC 
147, NGC 205, and NGC 221 are also shown.  The uncertainties due to distances 
are shown as dashed lines of slope unity while errors due to
distance-independent effects (e.g., photon statistics) are shown as the usual 
horizontal and vertical lines.  The solid lines, $L_{X,\ast}$ are the stellar 
X-ray contributions as determined from M31 and the hard + soft components of Cen A, 
while the dashed lines labeled $L_{SN}$ and $L_{grav}$ represent the energy 
released from supernovae ($L_{SN,1}$ from van den Bergh \& Tammann 1991, 
$L_{SN,2}$ from Turatto et al.\  1994) or available from thermalization 
and gravitational infall. \label{fig:lxlb} }
\end{figure}

There can be a large correction to the flux from 0.1--0.5 keV due to Galactic 
absorption, so we will limit our discussion to the luminosities determined for 
the 0.5--2.0 keV band (Table \ref{tab:xray}) which is fairly insensitive to 
Galactic absorption corrections. A logarithmic plot of $L_B$ (derived from 
Faber et al. \markcite{fab89}1989 magnitudes) against $L_X$ (Figure 
\ref{fig:lxlb}) for the sample is reproduced from Brown \& Bregman (\markcite{bro98}1998) with 
the addition of three dwarf galaxies: NGC 147, NGC 205, and NGC 221 (data 
processed according to the methods described in \S\ref{sec:sample}.  The galaxy 
NGC 221 (M 32) is a detection of a single point source, with a calculated X-ray 
luminosity (log$L_X = 37.39 \, {\rm ergs \, s}^{-1}$) in good agreement with 
Burstein et al. (\markcite{bur97}1997), while NGC 147 and NGC 205 are 
upper limits.  

{\small
\begin{table}[p]
\begin{center}
\caption{X-Ray Galaxy Properties \label{tab:xray}}
\newcommand{\erra}[1]{$\scriptstyle \pm{#1}$}
\newcommand{\errb}[2]{$\stackrel{+#1}{\scriptstyle -#2}$}

%\begin{tabular}{lrrcclrrc}
\begin{tabular}{ l@{\hspace{0.5in}}c@{\hspace{0.5in}}c@{\hspace{0.5in}}c }
\tableline \tableline

{} & 0.1--2.0 keV & \multicolumn{2}{c}{0.5--2.0 keV} \\
Name & log$L_X$ & log$L_X$ & log$\frac{L_X}{L_B}$ \\
{} & (erg s$^{-1}$) & (erg s$^{-1}$) & ($\frac{erg s^{-1}}{L_{\odot}}$)
\\ \tableline

N 0720 & 41.25 \erra{0.02} & 41.10 \errb{0.01}{0.02} & 30.15 \erra{0.06} \\
N 1316 & 41.25 \erra{0.01} & 41.08 \erra{0.01} & 29.74 \erra{0.06} \\
N 1344 & 39.75 \errb{0.14}{0.20} & 39.47 \errb{0.14}{0.20} & 28.81 \errb{0.15}{0.21} \\
N 1395 & 41.20 \erra{0.02} & 41.04 \erra{0.02} & 30.02 \erra{0.06} \\
N 1399 & 41.62 \erra{0.02} & 41.44 \erra{0.02} & 30.56 \erra{0.06} \\
N 1404 & 41.40 \erra{0.02} & 41.27 \erra{0.02} & 30.53 \erra{0.06} \\
N 1407 & 41.51 \erra{0.01} & 41.34 \erra{0.01} & 30.18 \erra{0.12} \\
N 1549 & 40.47 \errb{0.03}{0.04} & 40.04 \erra{0.03} & 29.31 \erra{0.07} \\
N 2768 & 40.57 \erra{0.06} & 40.41 \erra{0.06} & 29.62 \erra{0.13} \\
N 3115 & 40.07 \errb{0.06}{0.07} & 39.74 \errb{0.06}{0.07} & 28.91 \errb{0.08}{0.09} \\
N 3377 & 39.96 \errb{0.11}{0.15} & 39.42 \errb{0.11}{0.15} & 29.21 \errb{0.16}{0.19} \\
N 3379 & 39.94 \errb{0.15}{0.22} & 39.78 \errb{0.15}{0.22} & 29.29 \errb{0.16}{0.23} \\
N 3557 & 40.77 \erra{0.05} & 40.61 \erra{0.05} & 29.51 \erra{0.08} \\
N 3585 & 39.97 \errb{0.09}{0.12} & 39.84 \errb{0.09}{0.12} & 29.12 \errb{0.11}{0.13} \\
N 3607 & 40.97 \erra{0.02} & 40.82 \erra{0.02} & 29.64 \erra{0.12} \\
N 3923 & 41.02 \erra{0.02} & 40.90 \erra{0.02} & 29.91 \erra{0.06} \\
N 4125 & 41.24 \erra{0.03} & 41.01 \erra{0.03} & 29.85 \erra{0.12} \\
N 4278 & 40.88 \errb{0.04} {0.05} & 40.55 \errb{0.04}{0.05} & 29.83 \errb{0.07}{0.08} \\
N 4365 & 40.75 \erra{0.03} & 40.48 \erra{0.03} & 29.69 \erra{0.07} \\
N 4374 & 41.25 \erra{0.02} & 41.09 \erra{0.02} & 30.10 \erra{0.06} \\
N 4406 & 41.96 \erra{0.01} & 41.80 \erra{0.01} & 30.70 \erra{0.06} \\
N 4472 & 41.93 \erra{0.00} & 41.77 \erra{0.00} & 30.45 \erra{0.06} \\
N 4494 & 39.49 \errb{0.14}{0.22} & 39.28 \errb{0.14}{0.22} & 29.08 \errb{0.15}{0.23} \\
N 4552 & 41.09 \erra{0.01} & 40.92 \erra{0.01} & 30.21 \erra{0.06} \\
N 4621 & 39.91 \errb{0.08}{0.10} & 39.79 \errb{0.08}{0.10} & 29.01 \errb{0.14}{0.16} \\
N 4636 & 41.95 \erra{0.01} & 41.81 \erra{0.01} & 30.85 \erra{0.06} \\
N 4649 & 41.64 \erra{0.01} & 41.48 \erra{0.01} & 30.34 \erra{0.06} \\
N 4697 & 40.41 \erra{0.02} & 40.13 \erra{0.02} & 29.55 \erra{0.06} \\
N 5061 & 39.71 \errb{0.12}{0.16} & 39.54 \errb{0.12}{0.16} & 29.01 \errb{0.13}{0.17} \\
N 5102 & 37.82 \errb{0.12}{0.17} & 37.70 \errb{0.12}{0.17} & 28.75 \errb{0.17}{0.21} \\
N 5322 & 40.48 \erra{0.04} &40.11 \erra{0.04} & 29.31 \erra{0.13} \\
N 5846 & 42.15 \erra{0.01} & 42.01 \erra{0.01} & 30.75 \erra{0.12} \\
I 1459 & 41.36 \erra{0.01} & 41.19 \erra{0.01} & 30.05 \erra{0.06} \\
N 7507 & 40.25 \errb{0.19}{0.33} & 40.13 \errb{0.19}{0.33} & 29.31 \errb{0.20}{0.34} \\ \tableline

\end{tabular}
\end{center}
\end{table}
}

The X-ray-fainter galaxies appear to follow a linear relation 
(Figure \ref{fig:lxlb}) that can be compared to a stellar contribution 
derived from hard and soft X-ray components of Centaurus A 
(log $L_X/L_B = 28.96 \, {\rm ergs \, s}^{-1}/L_{\odot}$; Brown \& 
Bregman \markcite{bro98}1998, and references therein).  A stellar component may also be 
scaled from the M31 bulge, since it might be expected that the bulges 
of spiral galaxies exhibit the same spectral signatures as early-type galaxies 
dominated by stellar emission.  The M31 contribution (log $L_X/L_B = 29.21 \, {\rm ergs \, s}^{-1}/L_{\odot}$ and 
log $L_X = 38.93  \, {\rm ergs \, s}^{-1}$, Irwin \& Sarazin 
\markcite{irw98}1998) lies 0.3 dex higher than the Cen A stellar line, from 
which we infer that the stellar X-ray to optical luminosity ratio may not be a 
constant for all early-type systems. 

The X-ray luminosities of the brightest galaxies in the sample can be
compared to the maximum amount of energy produced by stellar motions and 
gravitational infall (log$L_{grav}$ = 23.57 + (5/3)log$L_B$, Brown \&
Bregman \markcite{bro98}1998).  This energy
depends partly upon the shape of the potential well which is measured by the stellar 
velocity dispersion.  The assumption is made that the thermalization of stellar mass 
loss is 100\% efficient. Figure \ref{fig:lxlb} also indicates that 
if the supernova rates of van den Bergh \& Tammann (\markcite{van91}1991,
upper line) are correct, supernovae can also provide energy sufficient to produced 
the high X-ray luminosities observed.  The lower (by a factor of 4.7) supernova 
energy line (log$L_{SN,2}$ = log$L_B$ + 30.22) was determined using a 
supernova rate given by Turatto, Capellaro, \& Benetti (\markcite{tur94}1994) 
and assumes a SNe energy of $10^{51}$ ergs. 

To determine the correlation between $L_X$ and $L_B$, we implemented
the ordinary least-squares (OLS) linear regression bisector method of 
Feigelson \& Babu (\markcite{fei92}1992) who discuss the applicability and 
effectiveness of several unweighted least-squares linear regression models to 
astronomical problems.  The usual method employed by astronomers is the least 
squares Y-on-X (Y/X) fit, which minimizes residuals in Y.  The OLS(Y/X) method 
is clearly preferred if it is known that one variable physically depends on another, 
or if the goal is to predict Y ($L_X$) given X ($L_B$).  The goal of this study is 
not predicting $L_X$ given $L_B$, but understanding the fundamental relationship 
between $L_X$ and $L_B$.  In this case, the assignment of a dependent or 
independent variable is uncertain, and so a symmetrical linear regression method is 
most appropriate as it is invariant to a change of variables (Isobe et al. \markcite{iso90}1990).  
Of the four OLS lines reviewed of this sort, the
OLS bisector (the line bisecting the OLS(Y/X) and (X/Y) lines) is
recommended.

The OLS bisector method yields a slope of 2.72$\pm$0.27 for our data 
(Table \ref{tab:xfit}), which is slightly steeper than previously reported 
due to the application of a resampling procedure recommended for small
samples (Feigelson \& Babu \markcite{fei92}1992). In Table \ref{tab:xfit}, we 
note that OLS(Y/X) fitting yields a flatter slope consistent with 
$m \approx$ 2.0--2.3 (Eskridge et al. \markcite{esk95}1995, White \&
Davis \markcite{whi97}1997). There is an increase from 2.7 to 2.9 in the 
$L_X$--$L_B$ slope if the background is chosen far from the galaxy center (see 
Table \ref{tab:xfit}). The galaxy NGC 5102 is excluded from all fits primarily 
because of it's very low $L_X/L_B$ ratio, which places it among the dwarf 
galaxies on the $L_X$--$L_B$ plot (Figure \ref{fig:lxlb}). NGC 5102 has a very 
blue integrated color of ($B - V$)$^0_T$ = 0.58 (de Vaucouleurs et al.
\markcite{vau76}1976), and a low metallicity 
which suggests that this galaxy recently had a starburst episode, and its X-ray 
emission is consistent with that of stars 
(van Woerden et al. \markcite{woe93}1993).

An estimate of the hot gas contribution to the X-ray emission can be derived
by subtracting a linear stellar component ($L_{X,\ast}$) from the data. The 
removal of the Cen A estimate of $L_{X,\ast}$ yields a slope 
of 3.32$\pm$0.46 (bisector, Table \ref{tab:xfit}) for the $L_X$--$L_B$ 
distribution. The bisector slope in this case was calculated using
the methods of Isobe et al. (\markcite{iso90}1990) since the distributed 
software of Feigelson \& Babu (\markcite{fei92}1992) does not accommodate 
upper limits. A gas component was not derived from M31 as we determined that 
the $L_X$ magnitude of the M31 stellar component was greater than the lowest 
luminosity galaxies (NGC 1344, NGC 3115, and NGC 5102) at the 3$\sigma$ level. 
Also, CO is abundant in the bulge of M31, which distinguishes it from 
early-type galaxies (Loinard, Allen, \& Lequeux \markcite{loi95}1995), and it has 
been suggested that the M31 bulge represents a post-starburst stage 
(Rieke, Lebofsky, \& Walker \markcite{rie88}1988). 

{\small
\begin{table}[t]
\begin{center}
\caption{Least-Squares (OLS) Fits \label{tab:xfit}}

\begin{tabular}{lrrcrr}
\tableline \tableline

{} & \multicolumn{2}{c}{Bisector} &  &  \multicolumn{2}{c}{Y/X}\\
{} & Intercept & Slope & & Intercept & Slope \\ \tableline

$L_X$-$L_B$,        & 11.06$\pm$2.92 & 2.72$\pm$0.27 & & 16.57$\pm$2.49 & 2.22$\pm$0.23 \\
no stellar subtraction      &                &               & &                &               \\
$L_X$-$L_B$,        & 8.75$\pm$3.95 & 2.94$\pm$0.36 & & 17.20$\pm$3.98 & 2.16$\pm$0.36 \\
no subtr.; alt. bkg.  \tablenotemark{a}    &                &               & &          & \\
$L_X$-$L_B$,        & 4.44$\pm$4.87  & 3.32$\pm$0.46 & & 10.82$\pm$4.24 & 2.73$\pm$0.39 \\
Cen A subtraction \tablenotemark{b}  &               & & &              &               \\
{ } & & & & & \\
$T_X$-$T_{\sigma}$, & 0.26$\pm$0.14  & 1.43$\pm$0.34 & & 0.04 $\pm$0.13 & 0.90$\pm$0.34 \\
B\&B sample         &                &               & &                &               \\
$T_X$-$T_{\sigma}$, & 0.20$\pm$0.04  & 0.90$\pm$0.10 & & 0.12 $\pm$0.05 & 0.70$\pm$0.12 \\
Davis \& White      &                &               & &                &               \\

\end{tabular}
\end{center}
%\tablenotetext{}{
Fits performed in log-space, utilizing the bootstrap 
resampling procedure recommended for small samples. $L_B$ in solar units.
NGC 5102 not included in fits. \\
%\tablenotetext{a}{
$^a$Background taken far from galaxy center. Galaxies NGC 1344, 
NGC 3377, NGC 4494, NGC 5061, NGC 5102, \& NGC 7507 not included. \\
%\tablenotetext{b}{
$^b$Hard plus soft X-ray components subtracted.
\end{table}
}

\subsection{$T_X$--$T\sigma$ Correlation}
\label{sec:txtsig}

Brown \& Bregman (\markcite{bro98}1998) confirm a correlation between 
the fitted X-ray gas temperature ($T_X$) and the stellar velocity dispersion 
temperature ($T_{\sigma}$), in addition to the $L_X$--$L_B$ relationship. The 
slope of the log$T_X$--log$T_{\sigma}$ relationship is found to be 
1.43$\pm$0.34 (see Table \ref{tab:xfit}), which is slightly steeper than a slope of unity that is
expected in cooling flow models.  Also not theoretically expected, is a large dispersion in $T_X$ about the best fit line.

\begin{figure}
%\figurenum{3}
\plotone{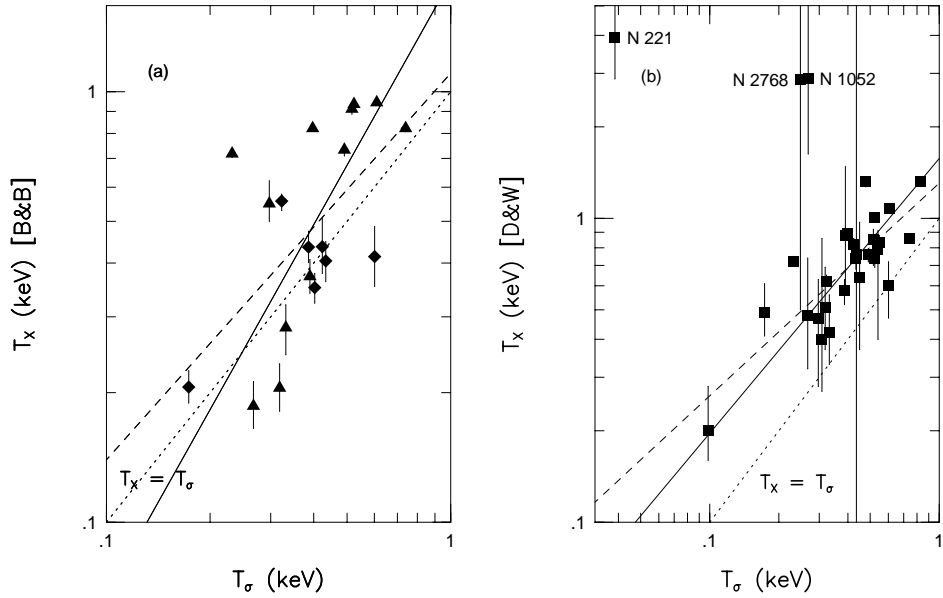}
\caption{ The stellar velocity dispersion temperature 
($T_{\sigma}$) vs. the fitted gas temperature for nineteen galaxies in our 
sample (a), and the Davis \& White sample (b). In both plots, the OLS bisector 
fit is represented by the solid line and the OLS(Y/X) fit is represented by the 
dashed line. The dotted line denotes the $T_X$ = $T_{\sigma}$ relation. The 
errors in (a) are at the 90\% confidence level. \label{fig:xtemp}}
\end{figure}

The slope to the temperature data differs from that reported by Davis \& White 
(\markcite{dav96}1996) who published temperatures for 30 galaxies.
The causes of this difference are discussed briefly in Brown \& Bregman 
(\markcite{bro98}1998). One cause lies in the methods these two groups
use in fitting the data. Our temperature data is reproduced in 
Figure \ref{fig:xtemp}a with the addition of a Y-on-X fitted line, and
compared to the data of Davis \& White (\markcite{dav96}1996, Figure \ref{fig:xtemp}b) in
illustration of the difference in statistical methods. For the Davis \& White 
(\markcite{dav96}1996) sample, we plot $T_{\sigma}$ instead of $\sigma$,
where $T_{\sigma}$ is derived from Faber et al. (\markcite{fab89}1989)
and Dressler et al. (\markcite{dre91}1991) velocity dispersions.  As 
stated in \S\ref{sec:lxlb}, the regression method utilized is a function of the 
problem being addressed.

Davis \& White (\markcite{dav96}1996) find gas temperatures everywhere hotter than the expected velocity dispersion temperature, which is a striking difference between their data and ours. Below $T_{\sigma} \approx 0.45$ keV we
find a more or less symmetrical distribution of gas temperatures about the
$T_X$ = $T_{\sigma}$ line.  In cases where we fit $T_X > 0.5$ keV, we
find that our temperatures agree well with Davis \& White (\markcite{dav96}1996) 
for the same galaxies. Below $T_X = 0.5$ keV, our temperatures range from 
$\sim$25--60\% below the values of Davis \& White (\markcite{dav96}1996). This difference cannot simply be the result of the 
number of temperature components allowed (see Brown \& Bregman 
\markcite{bro98}1998), as three of
our ``low" $T_X$ galaxies were fit with single-temperatures. Davis \& White 
(\markcite{dav96}1996) report extremely low solar abundances
(Z $\leq 0.06$ solar) for the three galaxies by allowing Z to be a fit parameter,
and at low abundances, derived temperatures become higher.  

A discussion of our data as compared to {\em ASCA} derived temperatures
is given in Brown \& Bregman (\markcite{bro98}1998). Our temperatures are
presented in Table \ref{tab:tfit} along with those of Davis \& White (1996), and 
Buote \& Fabian (\markcite{buo98}1998) {\em ASCA} temperatures for comparison.

{\small
\begin{table}[p]
\begin{center}
\footnotesize
\caption{Gas Temperatures \label{tab:tfit}}

\newcommand{\na}{\phm{\tablenotemark{a}}}
\newcommand{\no}{\nodata\phm{*}}
\newcommand{\erra}[1]{$\scriptstyle \pm{#1}$}
\newcommand{\errb}[2]{$\stackrel{+#1}{\scriptstyle -#2}$}

\begin{tabular}{lclllclclll}
\tableline \tableline

Name & $T_{\sigma}$ & $T_X$ & $T_X$ & $T_X$ &  &
Name & $T_{\sigma}$ & $T_X$ & $T_X$ & $T_X$ \\
{} &  & B\&B & D\&W & B\&F &  &
 &  & B\&B & D\&W & B\&F \\ \tableline

N 0221 & 0.039 \na & \no & 3.94 \errb{1.73}{1.06} & \nodata    & & N 4261 & 0.549 \na & \no & 0.83 \errb{0.05}{0.06} & \nodata    \\
N 0720 & 0.387 \na &  0.436* \errb{0.039}{0.035} & 0.58 \errb{0.05}{0.06} & 0.63  & & N 4278 & 0.450 \na & \no & 0.64 \errb{0.33}{0.27} & \nodata    \\
N 1052 & 0.269 \na & \no & 2.88 \errb{5.92}{1.25} & \nodata    & & N 4365 & 0.390 \na &  \no & 0.88 \errb{0.60}{0.29} & \nodata   \\
N 1316 & 0.403 \na &  0.351* \erra{0.028} & \nodata & \nodata  & & N 4374 & 0.524 \na & \no & 0.74 \erra{0.05} & 0.70 \errb{0.05}{0.04} \\
N 1344 & 0.163 \na & \no & \nodata & \nodata  & & N 4406 & 0.397 \na & 0.823\phm{*} \errb{0.010}{0.009} & 0.89 \erra{0.01} & 0.73 \errb{0.03}{0.02} \\
N 1395 & 0.424 \na &  0.437* \errb{0.071}{0.058} & 0.82 \errb{0.04}{0.06} & \nodata     & & N 4472 & 0.524 \na &  0.936\phm{*} \errb{0.007}{0.006} & 1.01 \erra{0.02} & 0.97        \\
N 1399 & 0.610 \na &  0.944\phm{*} \erra{0.009} & 1.08 \errb{0.02}{0.01} & 1.27       & & N 4486 & 0.831 \na & \no & 1.32 \errb{0.03}{0.04} & \nodata    \\
N 1404 & 0.323 \na &  0.557* \errb{0.021}{0.027} & 0.62 \errb{0.03}{0.02} & 0.56        & & N 4494 & 0.300 \tablenotemark{a} & \no & 0.20 \errb{0.08}{0.04} & \nodata               \\
N 1407 & 0.517 \na &  0.913\phm{*} \errb{0.027}{0.025} & 0.85 \errb{0.07}{0.15} & 0.92  & & N 4552 & 0.434 \na &  0.405* \errb{0.044}{0.042} & 0.74 \errb{0.07}{0.06} & \nodata     \\
N 1549 & 0.267 \na &  0.186\phm{*} \errb{0.026}{0.021} & 0.48 \errb{0.26}{0.16} & \nodata  & & N 4621 & 0.367 \na & \no  & \nodata & \nodata                  \\
N 2768 & 0.248 \na & \no & 2.87 \errb{16.13}{2.37} & \nodata   & & N 4636 & 0.232 \na &  0.717\phm{*} \erra{0.014} & 0.72 \errb{0.03}{0.02} & 0.66         \\
N 3115 & 0.450 \na & \no & \nodata  & \nodata                  & & N 4649 & 0.740 \na &  0.823\phm{*} \erra{0.015} & 0.86 \errb{0.02}{0.01} & 0.83         \\
N 3377 & 0.108 \na & \no & \nodata & \nodata                   & & N 4696 & 0.478 \tablenotemark{b} & \no & 1.32 \errb{0.04}{0.05} & \nodata               \\
N 3379 & 0.257 \na & \no & \nodata & \nodata                   & & N 4697 & 0.173 \na &  0.206* \errb{0.019}{0.017} & 0.49 \errb{0.12}{0.08} & \nodata     \\
N 3557 & 0.541 \na & \no & 0.79 \errb{0.12}{0.39} & \nodata    & & N 5061 & 0.233 \na & \no & \nodata & \nodata                   \\
N 3585 & 0.308 \na & \no & 0.40 \errb{0.46}{0.13} & \nodata    & & N 5102 & 0.500 \tablenotemark{a} & \no & \nodata & \nodata     \\
N 3607 & 0.390 \na &  0.372\phm{*} \errb{0.035}{0.032} & \nodata & \nodata  & & N 5322 & 0.319 \na &  0.205\phm{*} \errb{0.028}{0.024} & 0.51 \errb{0.18}{0.14} & \nodata  \\
N 3923 & 0.297 \na &  0.549\phm{*} \errb{0.071}{0.049} & 0.47 \errb{0.16}{0.19} & 0.64     & & N 5846 & 0.491 \na &  0.733\phm{*} \errb{0.024}{0.023} & 0.76 \errb{0.03}{0.04} & 0.67 \errb{0.04}{0.03} \\
N 4105 & 0.436 \tablenotemark{b} & \no & 0.76 \errb{4.40}{0.69} & \nodata   
& & I 1459 & 0.602  \na &  0.414* \errb{0.072}{0.061} & 0.60 \errb{0.12}{0.13} & 2.69        \\
N 4125 & 0.332 \na &  0.283\phm{*} \errb{0.037}{0.038} & 0.42 \errb{0.14}{0.09} & \nodata  & & N 7507 & 0.361 \na & \no  & \nodata & \nodata                  \\ \tableline

\end{tabular}
\end{center}
%\tablenotetext{}{
B\&B - Brown \& Bregman (1998). D\&W - Davis \& White (1996).
B\&F - Buote \& Fabian (1998). Only B\&F galaxies concurrent with B\&B listed.
All temperatures given in keV. $T_{\sigma}$ derived 
from Faber et al. (1989) values. $H_0$ = 50 km/s/Mpc used throughout.  
Starred values in the $T_X$ B\&B column denote a two temperature fit value 
where the 2nd component is fixed at 2.0 keV. B\&B errors at the 90\% confidence level. \\
%\tablenotetext{a}{
$^a$Adopted value for $T_{\sigma}$. \\
%\tablenotetext{b}{
$^b T_{\sigma}$ from Dressler et al. (1991).
\end{table}
}

\subsection{The galaxy environment}
\label{sec:environment}

We suggested in our previous paper that the observed X-ray luminosity of a
galaxy was strongly influenced by its environment because the most X-ray
luminous galaxies were in clusters or in the centers of groups. Here, we
examine this by quantifying the environment richness through the use of
the local galaxy density, $\rho$.  We determine which objects are X-ray
luminous for their $L_B$ by comparing $L_X/L_B$ to the individual $\rho$ 
of each galaxy in our sample. 

\begin{figure}
%\figurenum{4}
\plotone{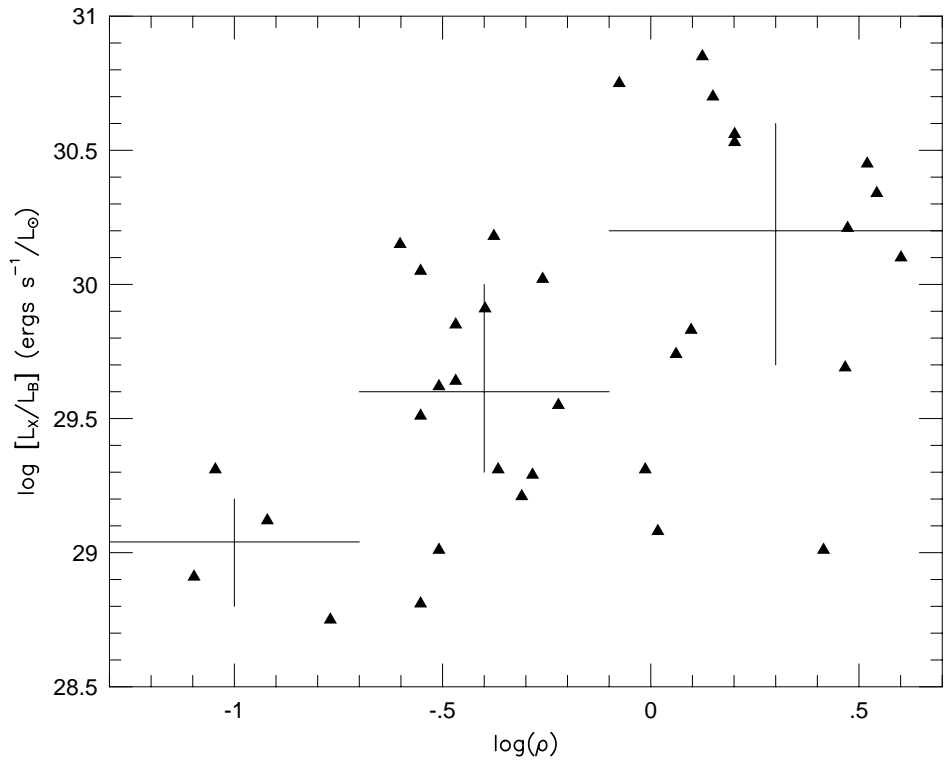}
\caption{ The 
luminosity ratio ($L_X/L_B$) vs. the local density of galaxies brighter than 
-16 mag in the vicinity of each galaxy (Tully 1988 values).  The horizontal 
lines delineate the width of each bin and mark the median ration within each 
bin. The vertical lines are the magnitude of the upper and lower quartile 
values within each bin. \label{fig:dens}}
\end{figure}

In Figure \ref{fig:dens}, the ratio of the X-ray-to-optical luminosity is 
plotted against the Tully (\markcite{tul88}1988) local density of galaxies ($\rho$) 
brighter than -16 mag in the vicinity of each sample galaxy.  The galaxy density 
is calculated such that an isolated galaxy will have a local density of 
$\rho$ = 0.06 galaxies Mpc$^{-3}$ by virtue of its own presence.  The local 
density in a richer environment, for example the center of the Virgo cluster, 
would be approximately 5 galaxies Mpc$^{-3}$. For our sample, no high luminosity 
systems are found in the most isolated environments ($\rho <$ 0.2 galaxies 
Mpc$^{-3}$), where the median log $L_X/L_B$ is plotted on Figure \ref{fig:dens} 
at $\approx 29.1 \, {\rm ergs \, s}^{-1}/L_{\odot}$.  The upper and 
lower 25\% quartile values for log $L_X/L_B$ in this region are 29.2 and 28.8 
respectively (both in $\, {\rm ergs \, s}^{-1}/L_{\odot}$).  The highest luminosity galaxies 
are only found in the densest environments ($\rho >$ 0.79 galaxies Mpc$^{-3}$), 
however lower luminosity ellipticals are also found in this region contributing 
to a collective median luminosity ratio of $\approx 30.2 \, {\rm ergs \, s}^{-1}/L_{\odot}$, 
with log $L_X/L_B = 30.6  \, {\rm ergs \, s}^{-1}/L_{\odot}$ at the
upper 25\% quartile and 29.7 $\, {\rm ergs \, s}^{-1}/L_{\odot}$ at the lower 
25\% quartile. For 0.2 $< \rho <$ 0.8 galaxies Mpc$^{-3}$, the median luminosity ratio has a moderate 
value of log($L_X/L_B$) $\approx 29.6 \, {\rm ergs \, s}^{-1}/L_{\odot}$ and 
upper and lower 25\% quartile log $L_X/L_B$ values of 29.3 and 30.0 respectively  
(in $\, {\rm ergs \, s}^{-1}/L_{\odot}$).  

The data appear to follow a correlation albeit with broad dispersion, and
there is a lower limit to the luminosity ratio for $-1.2 <$ log $\rho <$ 0.6, 
with no galaxies found below log $L_X/L_B = 28.8 \, {\rm ergs \, s}^{-1}/L_{\odot}$. 
We applied statistical tests to the data to quantify the apparent correlation. 
Three non-parametric tests for bivariate data, Kendall's Tau, Spearman's Rho, 
and Cox Proportional Hazard, were performed, each of which determined a 
correlation in the data at better than the 99.7\% confidence level.

\begin{figure}
%\figurenum{5}
\plotone{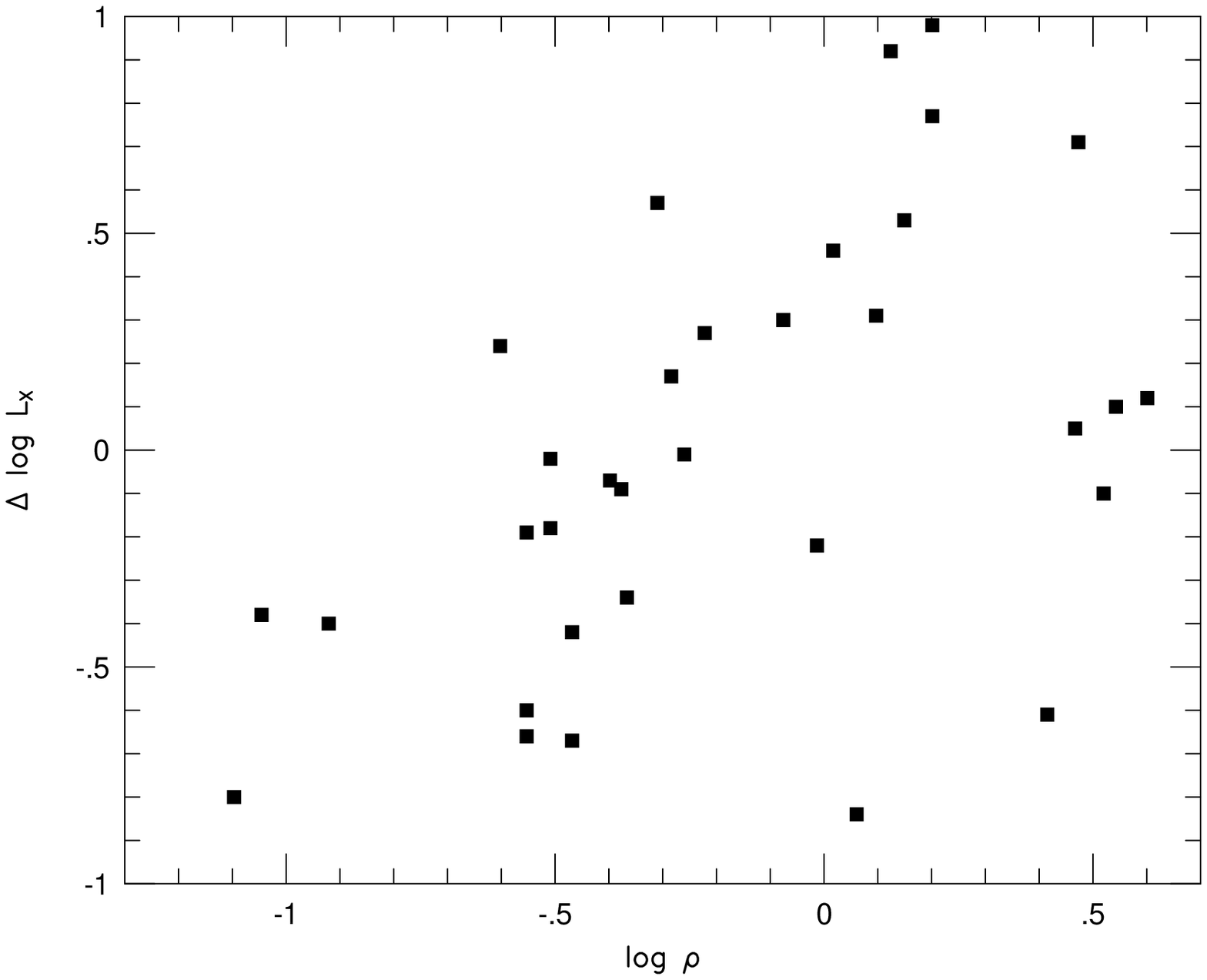}
\caption{The residuals in $L_X$ vs. the local density of 
galaxies brighter than -16 mag in the vicinity of each galaxy (Tully 1988 
values).  The residuals in $L_X$ are relative to the least squares fit between
$L_X$ and $L_B$. Error bars have been repressed in this plot. \label{fig:resid}}
\end{figure}

We additionally examined the $L_X$ residuals for the OLS(Y/X) fit,
excluding NGC 5102, as they
might relate to the galaxy environment (see Figure \ref{fig:resid}).  
The residuals are defined to be the difference between $L_X$ and the OLS
bisector fit to $L_X$ and $L_B$.  The Kendall's Tau test
indicates a correlation exists at $>$ 99\% 
confidence.  The correlation implies that the galaxies brightest for their 
optical luminosity (i.e, NGC 1399, NGC 4636, NGC 4552) are found in the 
spatially densest regions (relative to the rest of the sample). Occasionally, a
galaxy with a low $L_X$, given $L_B$, is found in a dense environment
indicating that complex mechanisms affect observed X-ray luminosities.

\begin{figure}
%\figurenum{6}
\plotone{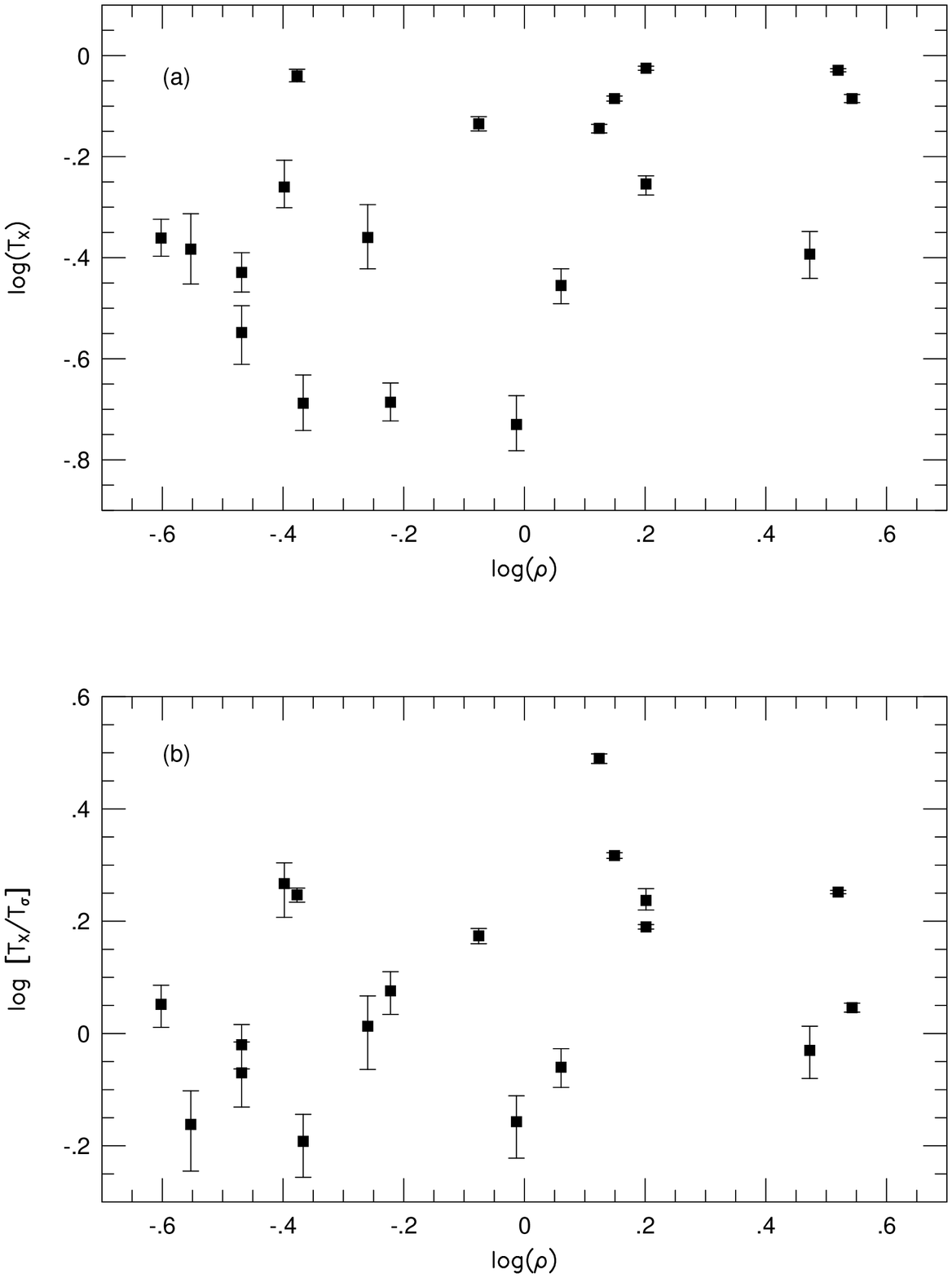}
\caption{The gas temperature ($T_X$) vs. the local density 
of galaxies brighter than -16 mag in the vicinity of each galaxy (Tully 1988 
values, 6a), and the temperature ratio ($T_X/T_{\sigma}$ vs. $\rho$, 6b).  The errors 
in $T_X$ are at the 90\% confidence level. \label{fig:t_dens}}
\end{figure}

The fitted X-ray temperatures, as a function of spatial density, exhibit a
weak correlation at the 92-95\% confidence level, depending upon if one uses the
Kendall's Tau test, the Spearman's Rho test, or the Cox Proportional Hazard
model (Figure \ref{fig:t_dens}a).  The hottest galaxies (k$T_X \sim 0.8$ keV)
are found in a range of environments 
as are galaxies with k$T_X \sim 0.45$ keV. The ratio of $T_X$ to $T_{\sigma}$
as a function of density exhibits an even weaker correlation, as shown in Figure
\ref{fig:t_dens}b.

\subsection{Malmquist bias}

The issue of bias is an important one for our sample, so we have addressed
the statistical and geometrical effects that are often discussed as 
Malmquist bias. One effect is of concern whenever measured distances
contain uncertainty. Along a line of sight, as one goes farther in distance, the
volume of space associated with a given solid angle increases. This has the
effect of giving greater weight to larger distances. This is the ``classical" 
Malmquist bias, and is corrected for in the distance measurements used in our 
analyses (Faber et al. \markcite{fab89}1989).

The other type of Malmquist bias occurs in magnitude-limited samples and
reflects the fact that the intrinsic luminosity function is not being
equally or completely sampled at all distances.  For the most distant galaxies, 
only the high luminosity part of the luminosity function is sampled, whereas 
for the nearest galaxies, much more of the luminosity function is sampled.  
Although this is an important issue for some investigations, such as those 
using standard candles as distance indicators (e.g., Teerikorpi 
\markcite{tee97}1997), it is not important for this study, which does not 
require that the Schecter luminosity function be fully sampled.  Our sample 
merely requires that the galaxies be representative of those that comprise the 
Schecter luminosity function, and that they were chosen in a fashion that does 
not bias the X-ray luminosity.  These galaxies where chosen independently of 
their X-ray properties, each was detected in X-rays, and they lie far above the
magnitude limit of the Faber et al. (1989) sample, so this should 
comprise an unbiased sample.

\section{Discussion and Conclusions}

\subsection{The Role of Environment}

One of results of our work is the demonstration that environment has a central
influence on the X-ray luminosity as seen in the positive correlation of the
$L_X/L_B$ ratio with galaxy density. In the lowest density environments,
only X-ray faint galaxies are found (low $L_X/L_B$ ratios), and the
galaxies with the highest $L_X/L_B$ ratios are found in fairly dense
environments. However, galaxies in dense environments exhibit a wide range in
$L_X/L_B$, suggesting that environment does not have the same effect on all
galaxies, and we suggest a natural explanation for this.

The distribution of $L_X/L_B$ in Figure \ref{fig:dens} reveals that there is a positive
correlation between local environmental richness and X-ray brightness, but
with significant dispersion at moderate and high galaxy density. In an effort
to understand this distribution, it is important to recognize that environment
can have both positive and negative effects on the X-ray brightness of a
galaxy (e.g., Takeda, Nulsen, \& Fabian \markcite{tak84}1984). Groups and
clusters with a significant ambient medium can lead to stripping of the
interstellar gas from a galaxy, leading to a substantial reduction in $L_X/L_B$.
However, for galaxies moving slowly through relatively cool clusters or
groups, the galaxy may be able to accrete this external material, increasing
$L_X/L_B$ (Brighenti \& Mathews \markcite{bri98}1998). Also, an ambient group or cluster medium could
stifle galactic winds, should they exist, causing the galaxy to retain its hot
gas locally, which also will increase $L_X/L_B$. Therefore, galaxies in
clusters or groups would have a range of $L_X/L_B$, where the low values
are determined by galaxies where stripping or winds occur, and the high values
represent systems that are accreting the ambient material or having their winds
stifled. In very rich clusters, like Coma, stripping is probably the dominant
process, but none of the galaxies in our sample lie in such a rich system. The
richest cluster in this sample is Virgo, where stripping is expected to occur
in some of the galaxies, but not most of them (Gaetz, Salpeter, \& Shaviv \markcite{gae87}1987).

The role of environment has been previously examined by White \& Sarazin
(\markcite{whi91}1991) using {\em Einstein Observatory} data to determine
if a correlation exists between local galaxy density and $L_X$. White \& 
Sarazin (1991) find the number of bright, neighboring galaxies within various 
projected distances from each X-ray galaxy, and perform linear regressions
of log $L_X$ against log $L_B$ and the local density of galaxies. They
conclude that the large dispersion in $L_X$ is not significantly reduced
by taking into account the individual local galaxy density, which may
be due to uncertainties in the numbers of neighbors found. White \&
Sarazin (1991) then binned the X-ray sample in order to reduce statistical
error. The X-ray sample is divided into four subsets: high-$L_X$
(log $L_X/L_B \gtrsim 30 \, {\rm ergs \, s}^{-1}/L_{\odot}$) detections, 
high-$L_X$ upper limits,
low-$L_X$ detections, and low-$L_X$ upper limits. The total number of
bright galaxies (again, within various projected distances) around
all galaxies in each subset is calculated yielding an averaged
number of bright galaxies per X-ray galaxy for each subgroup. 
They find that low-$L_X$ galaxies have $\sim 50$\% more neighbors
than high-$L_X$ galaxies, which the authors argue is expected if ram-pressure
stripping is a major factor in the $L_X$ dispersion.

The method of White \& Sarazin (\markcite{whi91}1991) is significantly 
different from ours.
Whereas they bin an incomplete sample of {\em Einstein Observatory} data and look at
bright neighbors in various projected distances, we use the tabulated
local galaxy density data of Tully (\markcite{tul88}1988) for each
galaxy in our complete sample. The Tully local density is calculated using a 
three-dimensional grid spaced at 0.5 Mpc. Also, our data includes
X-ray fainter galaxies, which allows us to extend the range of
X-ray-to-optical luminosity ratios. Our data extends down to log $L_X/L_B = 28.8 \, {\rm ergs \, s}^{-1}/L_{\odot}$ versus the Canizares, Fabbiano, \& Trinchieri 
\markcite{can87}(1987; used by White \& Sarazin \markcite{whi91}1991) data which extends down to log $L_X/L_B = 29.2 \, {\rm ergs \, s}^{-1}/L_{\odot}$. 
Since none of the galaxies in these samples lie in rich environments (i.e., $>$ 4 galaxies per Mpc$^3$), we cannot determine if ram-pressure stripping effects 
becomes significant in the observed X-ray luminosity in very dense environments.

\subsection{The Importance of Galactic Winds}

The extremely low values of $L_X/L_B$ seen in isolated galaxies is
probably the strongest evidence in favor of galactic winds. The X-ray
emission from these systems is so low that one is probably detecting the
X-ray contribution from stars rather than gas, indicating that these
galaxies are not gas-rich. For these galaxies, an ambient group or cluster
medium is not detected and they do not lie in a high velocity dispersion
group, so stripping is unlikely to occur. Furthermore, the rate of supernova
energy input is probably adequate to drive a galactic wind, as discussed
below. Therefore, the most viable mechanism for rendering the galaxies X-ray
weak is through galactic winds (see Pellegrini \& Ciotti \markcite{pel98}1998, and references therein).

Galactic winds will occur if the supernova heating rate is sufficient to raise
the temperature of the mass loss from stars above the escape temperature 
(from the galaxy center). A steady-state galactic wind will obey Bernoulli's law, 
which for a flow that has a small flow velocity when it reaches large radius is 
$\frac{5P}{2\rho}+\Phi=0$, where $\Phi$ is the gravitational potential and 
$\frac{5P}{2\rho}$ is the enthalpy. For a steady-state wind from the center of a 
galaxy where $\Phi(0)=8\sigma^{2}$,
\begin{equation}
T_{wind}=\frac{16}{5}\sigma^{2}\mu m_{p}/k
\end{equation}
The variable $\mu$ is the mean molecular weight, $m_{p}$ is the proton mass, and 
$k$ is the Boltzmann constant.
We use the supernova rate of Cappellaro et al. (\markcite{cap97}1997),
and energy per supernova of $10^{51}$ ergs, and the stellar mass loss rate
of Faber \& Gallagher (\markcite{fab76}1976) to determine gas
temperature entering the system:
\begin{equation}
T_{gas}=(\alpha_{SN}T_{SN}+\alpha_{\ast}T_{\ast})/(\alpha_{SN}+\alpha_{\ast})
\end{equation}
where $\alpha_{\ast}$ and $\alpha_{SN}$ are the mass loss rates from stars 
and supernovae (see, e.g., Mathews \& Brighenti \markcite{mab97}1997).

We then find that the most optically luminous galaxy that can sustain a total wind 
occurs at $L_{B}=5\times10^{10}L_{\odot}$. Less luminous galaxies 
can drive a wind, provided that high ambient pressure from the surrounding
environment does not prevent it. For galaxies more luminous than about 
$1.5\times10^{11}L_{\odot}$, the radius beyond which a wind can exist is far 
enough out that most of the stellar mass loss is retained by the galaxy (for galaxies 
with more complex models, partial winds may be more common; see Pelligrini \& 
Ciotti \markcite{pel98}1998). The largest uncertainty in estimating these values for 
$L_B$ is the energy released per supernova, which may be uncertain by about a 
factor of two.

Previously, we offered the suggestion that the galaxies with high $L_X/L_B$
values were X-ray bright because galactic winds were stifled (contained) by
a high-pressure ambient medium. This is energetically possible if the
supernova rate is correctly given by van den Bergh \& Tammann 
(\markcite{van91}1991), whose rate differs by about a factor of three from 
Cappellaro et al. (\markcite{cap97}1997; see Figure \ref{fig:lxlb}). 
The primary discrepancy between the two rate calculations is the correction for the 
inability to detect supernovae against the ambient light of the galaxy. A correction 
by a factor of three was suggested by van den Bergh \& Tammann 
(\markcite{van91}1991), while Cappellaro et al. (\markcite{cap97}1997) 
argue that the correction factor is only about 40\%. We calculated the distance
from the center of an elliptical galaxy within which a point source near the
magnitude limit of these surveys would fall below the $5\sigma$ detection
threshold, for a typical elliptical galaxy in these surveys. We find that only
5-30\% of the supernovae would have been missed due to this effect, similar to
the more accurate calculation of Cappellaro et al. (\markcite{cap97}1997). 

The supernova heating rate of Turatto, Capellaro, \& Benetti 
(\markcite{tur94}1994; line $L_{SN,2}$ in Figure \ref{fig:lxlb}) is insufficient to
account for the X-ray emission from the highest $L_X/L_B$ systems, which
are up to a factor of three above this line (an energy of $10^{51}$ erg per
supernova is used). Unless the energy per supernova has been underestimated by
a factor of three, we consider unlikely our original suggestion that stifled
winds can explain the highest $L_X/L_B$ systems. The source of energy in
these high luminosity systems is most likely gravitational, supplied by
accretion onto the system, as discussed by Mathews \& Brighenti (\markcite{mab98}1998) and Brighenti \& Mathews  (\markcite{bri99}1999).  
They show that the temperature and density distributions of hot gas in X-ray 
luminous galaxies, such as NGC 4472, are consistent with their calculations.

\subsection{An Overall Explanation of the $L_X$--$L_B$ Distribution}

Previously, there was hope that the cooling flow model could provide a
satisfactory explanation for the observed log$L_X$--log$L_B$ distribution. An
$L_X$--$L_B$ correlation with a slope of about 1.7 was predicted, which was
similar to observations, and the dispersion about this correlation was treated
as perturbations due (in part) to differences in galaxy evolution. The picture that 
we present is fundamentally different. We suggest that the steep observed 
correlation  is due to the transition from galaxies with total winds, to those with 
partial winds, to those where the gas is retained and can accrete additional material. 
As calculated above, we expect this transition from winds to retained gas to
occur in the range $10.7 < {\rm log}L_B (L_{\odot}) < 11.2$. Below 
log$L_B = 10.7 (L_{\odot}) $, one
only finds galaxies with small values of $L_X/L_B$, as would be expected for
systems with total winds. Also, all galaxies with log$L_B > 11.2$ have $L_X/L_B$
values well above that for purely stellar emission, consistent with retaining
their gas. Of equal importance is the influence of environment, which can
remove galactic gas through stripping, but can also help galaxies retain their
gas by stifling winds, and can add to the galactic gas content through
accretion. 

A critical difference between our picture and most others is that heating by 
supernovae can be comparable to, if not greater than the heating by the 
thermalization of stellar ejecta. This suggestion is not in complete agreement with 
other models or with some observations, the disagreement centering around the 
importance of winds and the metal enrichment introduced by the supernovae. 
Brighenti \& Mathews (\markcite{bri99}1999) study massive galaxies and 
use a supernova rate that is less than half the value determined by Cappellaro 
et al. (\markcite{cap97}1997).  With these supernova rates and galactic 
potential wells, winds are not important. They show that a range of nearly 
an order of magnitude in $L_X/L_B$ can be introduced by truncation of the 
galaxy as it interacts with neighbors (Brighenti \& Mathews \markcite{bri99}1999).

However, the observed range in $L_X/L_B$ is two orders of magnitude, and
when the stellar contribution is removed, the range in $L_X/L_B$ due to
the gas alone substantially exceeds two orders of magnitude.  It is unlikely
that truncation alone will produce the wide range in $L_X/L_B$.  Also, in
their model, one might expect that truncation would be least important
in regions of low galaxy density, leading to high values of $L_X/L_B$ in
such regions.  In conflict with this expectation, we find that galaxies in the lowest density environments only have low values of $L_X/L_B$.  The
great range of $L_X/L_B$ and the finding that fairly isolated galaxies 
have low values of $L_X/L_B$ is consistent with expectations of a model that
incorporates galactic winds.

A discussion of supernova rates also introduces an apparent conflict
between the expected and observed metallicity of the X-ray emitting gas
(see Brown \& Bregman \markcite{bro98}1998).  For the standard cooling
flow picture and the rates given by Cappellaro et al. (\markcite{cap97}1997), 
the metallicity should be several times solar
instead of near-solar (Loewenstein \& Mushotzky \markcite{loe97}1997; 
Buote \markcite{buo99}1999).
However, the metallicities are measured only for the most luminous
galaxies, and these are precisely the ones for which accretion of
circumgalactic gas dominates the gas content.  The metallicity prediction
must be revised downward due to the diluting effects of the accreted gas,
which may reduce the metallicity into the observed range.

\subsection{Predictions of the Models}

The various suggestions made by us and others has several immediate predictions
that will be tested by upcoming observations. First, the amount of the stellar
contribution to the X-ray emission should be spatially resolved by {\em Chandra},
leading to a clear determination of this value, and probably a determination
of the spectrum, separated from the spectrum of the hot gas. Another important
test will be the measurement of the rate of cooling gas, since it should be
low if supernovae cause galactic winds and approximately the stellar mass loss
rate if cooling flows are uniformly present. We predict that the low
$L_X/L_B$ galaxies will show little evidence of cooling gas, which will be
measured through the O VIII and O VII X-ray lines with {\em Chandra}, and through the
O VI line with {\em FUSE}. The high $L_X/L_B$ galaxies should have cooling rates
similar to the stellar mass loss rate, if not greater, although most models
agree on this prediction. Cooling flow models should show a significant
metallicity gradient from the outer to the inner parts, reflecting the stellar
metallicity gradient, and this effect should be even more pronounced for
galaxies that are accreting material from the surrounding group or cluster.
Finally, the ongoing optical supernova searches, being carried out with modern
CCD techniques, should yield an improved determination of the supernova rate.

\acknowledgements
     We would like to thank a variety of people for valuable
discussion: J. Irwin, J. Mohr, P. Hanlan, R. White,
M. Loewenstein, G. Worthey, J. Parriott, M. Roberts, D.
Hogg, and R. Mushotzky.  Special thanks is due to the members of
the {\em ROSAT} team and to the archiving efforts associated with the
mission.  Also, we wish to acknowledge the use of the NASA
Extragalactic Database (NED), operated by IPAC under contract with
NASA;  SLOPES, which implements the methods presented in Isobe et al.
(\markcite{iso90}1990), Babu \& Feigelson (\markcite{bab92}1992), and 
Feigelson \& Babu (\markcite{fei92}1992); 
ASURV (Isobe \& Feigelson \markcite{iaf90} 1990), which implements the 
methods presented in Isobe, Feigelson, \& Nelson (\markcite{ifn86} 1986). 
NASA has provided support for this work through grants
NAGW-2135, NAG5-1955, and NAG5-3247; BAB would like to acknowledge
support through a NASA Graduate Student Researchers Program grant
NGT-51408, and the support of the National Academy of Sciences postdoctoral associate program.

%\newpage


\begin{references}

\reference{bab92}Babu, G. J., \& Feigelson, E. D. 1992, Communications in 
   Statistics, Simulation \& Computation, 21, 533
\reference{bau96}Bauer, F., \& Bregman, J. N. 1996, \apj, 457, 382
\reference{bri98}Brighenti, F., \& Mathews, W. G. 1998, \apj, 495, 239
\reference{bri99}Brighenti, F., \& Mathews, W. G. 1999, \apj, 512, 65
\reference{bro98}Brown, B. A., \& Bregman, J. N. 1998, \apjl, 495, L75, astro-ph/9712209
\reference{buo99}Buote, D. A. 1999, \mnras, in press, astro-ph/9903278
\reference{buo98}Buote, D. A., \& Fabian, A. C. 1998, \mnras, 296, 977
\reference{bur97}Burstein, D., Jones, C., Forman, W., Marston, A. P., \&
  Marzke, R. O. 1997, \apjs, 111, 163
\reference{can87}Canizares, C. R., Fabbiano, G., \& Trinchieri, G. 1987, \apj, 
  312, 503
\reference{cap97}Cappellaro, E., Turatto, M., Tsvetkov, D.Y., Bartunov, O.S.,
  Pollas, C., Evans, R., \& Hamuy, M. 1997, \aap, 322, 431
\reference{dav91}David, L. P., Forman, W., \& Jones, C. 1991, \apj, 369 121
\reference{dav96}Davis, D. S., \& White, R. E. 1996, \apj, 470, L35
\reference{der89}D'Ercole, A., Renzini, A., Ciotti, L., \& Pellegrini, S. 1989,
  \apj, 341, L9
\reference{don90}Donnelly, R. H., Faber, S. M. \& O'Connell, R. M. 1990, \apj,
  354, 52
\reference{dre91}Dressler, A., Faber, S. M., \& Burstein, D. 1991, \apj,
  368, 54
\reference{esk95}Eskridge, P. B., Fabbiano, G., \& Kim, D.-W. 1995, \apjs,
  97, 141
\reference{fao89}Fabbiano, G. 1989, \araa, 27, 87
\reference{fab76}Faber, S. M., \& Gallagher, J. S. 1976, \apj, 204, 365
\reference{fab89}Faber, S. M., Wegner, G., Burstein, D., Davies, R. L., 
  Dressler, A., Lynden-Bell, D., \& Terlevich, R. J. 1989, \apjs, 69, 763
\reference{fei92}Feigelson, E. D., \& Babu, G. J. 1992, \apj, 397, 55
\reference{gae87}Gaetz, T. J., Salpeter, E. E., \& Shaviv, G. 1987, \apj, 316, 
  530
\reference{irw98}Irwin, J. A., \& Sarazin, C. L. 1998, \apjl, 494, 33
\reference{iaf90}Isobe, T., \& Feigelson, E. D. 1990, \baas, 22, 917
\reference{ifn86} Isobe, T., Feigelson, E. D., \& Nelson, P. I. 1986, \apj,
  306, 490
\reference{iso90}Isobe, T., Feigelson, E. D., Akritas, M. G., \& Babu, G. J.
  1990, \apj, 364, 104
\reference{loe87}Loewenstein, M., \& Mathews, W. G. 1987, \apj, 319, 614
\reference{loe91}Loewenstein, M., \& Mathews, W. G. 1991, \apj, 373, 445
\reference{loe97}Loewenstein, M., \&  Mushotzky, R. F. 1997, Proceedings of IAU 
  Symposium 187 on Cosmic Chemical Evolution
\reference{loi95}Loinard, L., Allen, R. J., \& Lequeux, J. 1995, \aap,
  301, 68
\reference{mab97}Mathews, W. G., \& Brighenti, F. 1997, \apj, 488, 595
\reference{mab98}Mathews, W. G., \& Brighenti, F. 1998, \apjl, 503, 15
\reference{mcm94}McMillan, R., Ciardullo, R., \& Jacoby, G. H. 1994, \aj,
  108, 1610
\reference{pel98}Pellegrini, S. \& Ciotti, L. 1998, \aap, 333, 433
\reference{rie88}Rieke, G. H., Lebofsky, M. J., \& Walker, C. E. 1988,
  \apj, 325, 679
\reference{tak84}Takeda, H., Nulsen, P. E. J., \& Fabian, A. C. 1984, \mnras, 
  208, 261
\reference{tee97}Teerikorpi, P. 1997, \araa, 35, 101
\reference{tul88}Tully, R. B. 1988, Nearby Galaxies Catalog (Cambridge
  University: Cambridge)
\reference{tur94}Turatto, M., Cappellaro, E., \& Benetti, S. 1994, \aj, 108,
  202
\reference{van91}van den Bergh, S., \& Tammann, G. A. 1991, \araa, 29, 363
\reference{woe93}van Woerden, H., van Driel, W., Braun, R., \& Rots, A. H. 1993
  \aap, 269, 15
\reference{vau76}de Vaucouleurs, G., de Vaucouleurs, A., and Corwin, H. G. 
  1976, {\em Second Reference Catalogue of Bright Galaxies} (RC2), (Austin:  
  Univ. of Texas Press)
\reference{ved88}Vedder, P. W., Trester, J. J., \& Canizares, C. R. 1988, 
  \apj, 332, 725
\reference{whi97}White, R. E. III, \& Davis, D. S. 1997, in Galactic and
  Cluster Cooling Flows, ed. N. Soker (ASP: San Francisco), p. 217
\reference{whi91}White, R. E. III, \& Sarazin, C. L. 1991, \apj, 367, 476
\end{references}
\end{document}